\documentclass[doublecol]{epl2}

\usepackage{graphicx}
\usepackage{bm}
\usepackage{hyperref}
\usepackage{subfigure}
\usepackage{amsmath,amssymb}
\usepackage{soul}
\begin{document}

\title{Time-invariant entanglement and sudden death of nonlocality for multipartite systems under collective dephasing}
\shorttitle{Time-invariant entanglement and sudden death of nonlocality}

\author{Mazhar Ali}
\institute{Department of Electrical Engineering, Faculty of Engineering, Islamic University Madinah, 107 Madinah, Saudi Arabia}

\abstract{ We investigate the dynamics of entanglement and nonlocality for multipartite quantum systems under collective dephasing. 
Using an exact and computable measure for genuine entanglement, we demonstrate the possibility of a non trivial phenomenon of time-invariant entanglement 
for multipartite quantum systems. We find that for four qubits, there exist quantum states, which are changing continously nevertheless their genuine 
entanglement remains constant. Based on our numerical results, we conjecture that there is no evidence of time-invariant entanglement for quantum states of 
three qubits. We point out that quantum states exhibiting time-invariant entanglement must live in both decoherence free subspace and in the subspaces 
orthogonal to it. The previous studies on this feature for two qubits can be recovered from our studies as a special case. 
We also study the nonlocality of quantum states under collective dephasing. We find that although genuine entanglement of quantum states may not change, 
however their nonlocality changes. We discuss the possibility of finite time end of genuine nonlocality.}


\maketitle

Quantum entanglement and nonlocality are features of quantum mechanics not only related to its foundation but also have applications in current 
and future technologies \cite{Horodecki-RMP-2009, gtreview,Brunner-RMP}. Due to growing efforts for an experimental realization of devices utilizing 
these features, it is essential to study the effects of noisy environments on quantum correlations. 
Such studies are an active area of research \cite{Aolita-review} and several authors have studied decoherence effects on quantum correlations for 
both bipartite and multipartite systems
\cite{Yu-work,lifetime,Aolita-PRL100-2008,bipartitedec,Band-PRA72-2005,lowerbounds,Lastra-PRA75-2007, Guehne-PRA78-2008, Lopez-PRL101-2008, 
Ali-work, Weinstein-PRA85-2012,Ali-JPB-2014}. 

One specific type of noise dominant in experiments on trapped atoms is caused by intensity fluctuations of electromagnetic fields which leads 
to collective dephasing process. The detrimental effects of collective dephasing noise on entanglement have been studied 
\cite{Yu-CD-2002, AJ-JMO-2007,Li-EPJD-2007, Karpat-PLA375-2011, Ali-PRA81-2010, Liu-arXiv}, however all these previous studies were 
restricted to a special orientation (z-axes) of the field. Recently, a more general approach has been worked out 
\cite{Carnio-PRL-2015, Carnio-NJP-2016}, where the authors addressed an arbitrary orientation of field. This general approach revealed an 
interesting feature of its dynamical process which is so called {\it freezing} dynamics of entanglement. It was shown that a specific two qubits 
state under certain orientation of the field may first decay upto some numerical value before suddenly stop decaying and 
maintain this stationary entanglement \cite{Carnio-PRL-2015}. 
Such behavior was also predicted for multipartite states. Recently, we have confirmed this freezing entanglement phenomenon for various genuinely 
entangled states of three and four qubits, including random states \cite{Ali-2016}. Another interesting dynamical feature under this type of 
decoherence is the possibility of completely {\it time-invariant} entanglement, however time-invariance phenomenon so far including this work has only been 
observed for a special orientation of field (z-axes). Time-invariant entanglement does not necessarily mean that the quantum states live in 
decoherence free subspaces (DFS). In fact the quantum states may change at every instance whereas their entanglement remain constant throughout the 
dynamical process. This feature was first observed for qubit-qutrit systems \cite{Karpat-PLA375-2011} and more recently for qubit-qubit 
systems \cite{Liu-arXiv}. In this Letter, we investigate the time-invariant phenomenon for genuine entanglement of multiqubit quantum systems. It is 
known that genuine entanglement is different than the entanglement among bipartition and this type of entanglement is only a peculiar feature of 
multipartite quantum states. We have looked for this phenomenon in Hilbert space of three qubits and our preliminary search suggests that 
it may not exist for this dimension of Hilbert space. 
However, we have explicitly observed this phenomenon for a family of quantum states of four qubits. The interesting difference between three 
and four qubit case is the fact that for three qubits, all off-diagonal elements of GHZ-diagonal states decay and there are no DFS for them, whereas for 
four qubits, there 
are some GHZ-diagonal states which live in DFS. We have detected this phenomenon by taking mixtures of GHZ states living in DFS and ones living in other 
orthogonal subspaces. The mixing probability for entangled state preserved in DFS must be larger than the probability of entangled state which decay, 
such that although combined states might change whereas their entanglement stay invariant. It is interesting that genuine entanglement also exibits 
time-invariance even though the entanglement among bipartition is not constant as evident by a change in the negative eigenvalues of the partially 
transposed matrix. Recent progress in the theory of multipartite entanglement has enabled us to study decoherence effects on actual multipartite genuine
entanglement and not on entanglement among bipartitions. In particular, the ability to compute genuine negativity for multipartite systems 
has eased this task \cite{Bastian-PRL106-2011}. 

Another concept related to non-classical correlations is quantum nonlocality. This feature says that the predictions made using quantum 
mechanics cannot be simulated by a local hidden variable model. The presence of nonlocal correlations can be detected via violation of some type of Bell 
inequalities \cite{Bell-Phys-1964}. The pure entangled states violate a Bell inequality, whereas mixed entangled states may not do 
so \cite{Gisin-Werner-1991}. However, all entangled states do exhibit some kind of hidden nonlocality \cite{Liang-PRA86-2011}. The well known 
Clauser-Horne-Shimony-Holt (CHSH) inequality \cite{CHSH-1969} for two qubits has been studied under decoherence both in 
theory \cite{Mazzola-PRA81-2010}, and experiment \cite{Xu-PRL-2010}. 
Several investigations of nonlocality of multipartite quantum states under decoherence have been carried out \cite{NLD}.
The extension of CHSH inequality for multipartite quantum systems 
has received considerable attention \cite{Mermin-PRL-1990, Ardehali-PRA-1992, Collins-PRL-2002, Bancal-PRL-2011}, however Svetlichny discovered 
the first method to detect genuine multipartite nonlocality \cite{Svetlichny-PRD-1987}. Violations of some of these inequalities in experiments 
have been reported \cite{Pan-expMBI, Bastian-PRL104-2010}. We have studied the effect of collective dephasing on genuine nonlocality of quantum states 
exhibiting time-invariant dynamics. We have found that these quantum states may loose their genuine nonlocality at a finite time. 

We review the basic notions of genuine multipartite entanglement and genuine nonlocality only for three parties $A$, $B$, and $C$. One can 
generalize these methods to more parties straightforwardly. 
A state is called separable with respect to some bipartition, say, $A|BC$, if it is a mixture of product states with respect 
to this partition, that is, 
$\rho = \sum_j \, p_j \, |\psi_A^j \rangle\langle \psi_A^j| \otimes |\psi_{BC}^j \rangle\langle \psi_{BC}^j|$, 
where $p_j$ form a probability distribution. We denote these states as $\rho_{A|BC}^{sep}$. 
Similarly, we can define separable states for the two other bipartitions, $\rho_{B|CA}^{sep}$ and $\rho_{C|AB}^{sep}$. 
Then a state is called biseparable if it can be written as a mixture of states which are separable with respect 
to different bipartitions, that is 
\begin{eqnarray}
 \rho^{bs} = \tilde{p}_1 \, \rho_{A|BC}^{sep} + \tilde{p}_2 \, \rho_{B|CA}^{sep} + \tilde{p}_3 \, \rho_{C|AB}^{sep}\,,
\end{eqnarray}
with $\tilde{p}_1 +\tilde{p}_2 +\tilde{p}_3 =1$.
Finally, a state is called genuinely multipartite entangled if it is not biseparable. In the rest of this paper, we always mean genuine
multipartite entanglement when we talk about entanglement. 

Genuine entanglement can be detected and characterized \cite{Bastian-PRL106-2011} by a technique based on positive partial transpose 
mixtures (PPT mixtures). A two-party state $\rho = \sum_{ijkl} \, \rho_{ij,kl} \, |i\rangle\langle j| \otimes |k\rangle\langle l|$ is PPT if 
its partially transposed matrix $\rho^{T_A} = \sum_{ijkl} \, \rho_{ji,kl} \, |i\rangle\langle j| \otimes |k\rangle\langle l|$ is positive semidefinite. 
The separable states are always PPT \cite{peresppt} and the set of separable states with respect to some partition 
is therefore contained in a larger set of states which has a positive partial transpose for that bipartition. 

Denoting PPT states with respect to fixed bipartition by $\rho_{A|BC}^{PPT}$, $\rho_{B|CA}^{PPT}$, 
and $\rho_{C|AB}^{PPT}$, we call a state as PPT-mixture if it can be written as 
\begin{eqnarray}
\rho^{PPTmix} = q_1 \, \rho_{A|BC}^{PPT} + q_2 \, \rho_{B|CA}^{PPT} + q_3 \, \rho_{C|AB}^{PPT}\,.
\end{eqnarray}
As any biseparable state is a PPT-mixture, therefore any state which is not a PPT-mixture is guaranteed to be genuinely multipartite entangled. 
The main advantage of considering PPT-mixtures instead 
of biseparable states comes from the fact that PPT-mixtures can be fully characterized by the method of semidefinite programming 
(SDP), a standard  method in convex optimization \cite{sdp}. Generally the set of PPT-mixtures is a very good approximation to the 
set of biseparable states and delivers the best known separability criteria for many cases; however, there are multipartite entangled 
states which are PPT-mixtures \cite{Bastian-PRL106-2011}. 
In order to quantify genuine multipartite entanglement, it was shown \cite{Bastian-PRL106-2011} that for the following optimization problem 
\begin{eqnarray}
\min {\rm Tr} (\mathcal{W} \rho)
\end{eqnarray}
with constraints that for all bipartition $M|\bar{M}$
\begin{eqnarray}
\mathcal{W} = P_M + Q_M^{T_M},
 \, \, \mbox{ with }
0 \leq P_M \leq \mathbb{I} \mbox{ and }
0 \leq Q_M  \leq \mathbb{I}\, 
\end{eqnarray}
the negative witness expectation value is multipartite entanglement monotone. 
The constraints just state that the considered operator $\mathcal{W}$ is a decomposable entanglement
witness for any bipartition. Since this is a semidefinite program, the minimum can be efficiently computed and the optimality of the solution can
be certified \cite{sdp}. We denote this measure by $E(\rho)$ or $E$-monotone in this paper. For bipartite systems, this monotone is 
equivalent to {\it negativity} \cite{Vidal-PRA65-2002}. For a system of qubits, this measure is bounded by  $E(\rho) \leq 1/2$ \cite{bastiangraph}.

For a brief description of genuine nonlocality, consider that each party can perform a measurement $X_j$ with result $a_j$ for $j = A,B,C$. The 
joint probability distribution $P(a_A a_B a_C|X_A X_B X_C)$ may exhibit different notions of nonlocality. It may be that it cannot be written 
in local form as
\begin{eqnarray}
 P(a_A a_B a_C|X_A X_B X_C) = \int d\lambda \, p_\lambda \, P_A(a_A|X_A \lambda) \nonumber \\ P_B(a_B|X_B \lambda)\, P_C(a_C|X_C \lambda) \,,
\label{Eq:PLV}
 \end{eqnarray}
where $\lambda$ is a shared local variable. Such nonlocality can be tested by standard Bell inequalities and it can not capture the 
genuine nonlocality. As an example consider that parties $A$ and $B$ are nonlocally correlated but uncorrelated from party $C$. It is still 
possible that $P$ cannot be written as Eq.(\ref{Eq:PLV}), although the system has no genuine tripartite nonlocality \cite{Bancal-PRL-2011}. 
Genuine nonlocality can be detected if one makes sure that $P$ cannot be written as 
\begin{eqnarray}
P_G(a_A a_B a_C|X_A X_B X_C) = \sum_{m = 1}^{3} \, p_m \, \int d\lambda \rho_{ij}(\lambda) \nonumber \\ 
P_{ij}(a_i a_j|X_i X_j \lambda) \, P_m(a_m|X_m \lambda)\, \,,
\label{Eq:GLV}
 \end{eqnarray}
that is, $P$ cannot be reproduced by local means even if any two of parties come together and act jointly to produce bipartite nonlocal 
correlations with probability distribution $\rho_{ij}(\lambda)$, where $ij$ denotes for all possible partitions. 
We focuss on the possibility that each party $j$ is allowed to two measurements  
$X_j$ and $X'_j$ with outcomes $a_j$ and $a'_j$ such that $a_j, \, a'_j \in \{-1, 1\}$. For an initial state 
$|GHZ\rangle = (|0000\rangle + |1111\rangle)/\sqrt{2}$, we consider the Ardehali inequality \cite{Ardehali-PRA-1992} 
$\langle B_A \rangle \leq 4$, where   
\begin{eqnarray}
\mathcal{B}_A =  \big( A_1 X_2 X_3 X_4 + B_1 X_2 X_3 X_4 - [A_1 X_2 Y_3 Y_4 + perm ] \nonumber \\ - [B_1 X_2 Y_3 Y_4 + perm] 
- [A_1 X_2 X_3 Y_4 + perm] \nonumber\\ + [B_1 X_2 X_3 Y_4 + perm] + A_1 Y_2 Y_3 Y_4 - B_1 Y_2 Y_3 Y_4 \big) \,, 
\end{eqnarray}
and the sum in square brackets include all distinct permutations on last three qubits, $A_1 = (X_1 + Y_1)/\sqrt{2}$, and $B_1 = (X_1 - Y_1)/\sqrt{2}$. 
The expectation value for GHZ state is $\langle B_A \rangle = 8 \sqrt{2}$, which is the maximum violation for four qubits.

We consider our qubits as atomic two-level systems with energy splitting $\hbar \omega$. The splitting is controlled by a homogeneous magetic field. 
The Hamiltonian for a single atom is given as 
\begin{eqnarray}
\hat{H}_\omega = \frac{\hbar \, \omega}{2} \, \vec{n} \cdot \vec{\sigma} \,,
\label{Eq:Hamil}
\end{eqnarray}
where $\vec{n} = n_x \hat{x} + n_y \hat{y} + n_z \hat{z}$ is the orientation of magnetic field and 
$\vec{\sigma} = \sigma_x \hat{x} + \sigma_y \hat{y} + \sigma_z \hat{z}$ is the vector of Pauli matrices. 
This time independent Hamiltonian generates the propagator 
\begin{eqnarray}
U_\omega(t) = {\rm e}^{- i \, H_\omega \, t/\hbar } = {\rm e}^{- i \, \omega \, t/2 \, \, \, {\bf n} \cdot {\bf \sigma} } \,. 
\label{Eq:Uw}  
\end{eqnarray}
We can introduce a pair of orthogonal projectors 
\begin{eqnarray}
\Lambda_\pm = \frac{I_2 \pm {\bf n}\cdot {\bf \sigma}}{2} \, ,
\label{Eq:TE}
\end{eqnarray}
to write the propagator in terms of them. Let us consider $N$ non-interacting atoms (qubits), so that the propagator for these collection of atoms 
can be written as \cite{Carnio-PRL-2015} 
\begin{eqnarray}
U_\omega(t)^{\otimes \, N} =& \, (e^{-i \, \omega \, t} \, \Lambda_+ \, + e^{i \, \omega \, t} \, \Lambda_-)^{\otimes \, N} \nonumber \\&
= \sum_{j = 0}^N \, e^{i\, \omega \,t (j-N/2) } \, \Theta_j \, ,
\label{Eq:UwN}
\end{eqnarray}
where the operators $\Theta_j$ are defined as 
\begin{eqnarray}
\Theta_j = \frac{1}{j! \, (N-j)!} \, \sum_{s \in \sum_N} \, V_s \, \big[ \Lambda_-^{\otimes \, j} 
\otimes \Lambda_+^{\otimes \, N-j} \big] \, V_s^\dagger \, , 
\label{Eq:Theta}
\end{eqnarray}
where $\sum_N$ represents the symmetric group and $V_s$ are the permutations in operator space of $N$ qubits. 

As there are fluctuations in the magetic field strength, the integration over it will induce a probability distribution $p(\omega)$ of 
characteristic energy splitting. Therefore the time evolution of the combined state of $N$ atoms can be written as \cite{Carnio-PRL-2015}
\begin{eqnarray}
\rho(t) = \int p(\omega) \, U_\omega(t)^{\otimes \, N} \, \rho(0) \, U^\dagger_\omega(t)^{\otimes \, N} \, d\omega \,.
\end{eqnarray}
In writing this equation, we have assumed that the field fluctuations occur on time scale which are longer than the time over which the combined 
state of $N$ atoms evolve under unitary propagator $U_\omega(t)^{\otimes N}$. Substituting the above derived format for the unitary propagator, we can 
write the time evolved state as 
\begin{eqnarray}
\rho(t) = \sum_{j,k = 0}^N \, M_{jk}(t) \, \Theta_j \, \rho(0) \, \Theta_k \, ,
\label{Eq:TES}
\end{eqnarray}
where $M_{jk}(t)$ are elements of the Toeplitz matrix $M(t)$, which can be obtained by the relation $M_{jk}(t) = \phi[(j-k)t]$, where $\phi(t)$ is 
the characteristic function of the probability distribution $p(\omega)$, defined as
\begin{eqnarray}
 \phi(t) = \int \, p(\omega) \, e^{i \, \omega \, t} \, d\omega \, .
\end{eqnarray}
It has been demonstrated that time evolution form Eq.(\ref{Eq:TES}) is both trace preserving and positivity preserving \cite{Carnio-PRL-2015}. 
In order to study the exact behavior of multipartite quantum states, 
it is convenient to obtain an exact expression for state $\rho(t)$, in terms of a spectral distribution $p(\omega)$ characterizing the fluctuations. 
As an example, we take the Lorentzian distribution also known as Cauchy distribution, defined as
\begin{equation}
 p(x) = \frac{\gamma^2}{\pi \, \gamma \, \big[ (x - x_0)^2 + \gamma^2 \big]}\,.
\end{equation}
For standard Cauchy distribution, the characteristic function turns out to be 
\begin{equation}
 \phi(t) = \int \, p(x) \, e^{i \, x \, t} \, dx = e^{-|t|} \,,
\end{equation}
here $t$ denotes dimensionless quantity usually taken as $ \Gamma \, t$. 
The time evolution of an arbitrary initial state can be obtained straightforwardly. In general there are no decoherence free subspaces (DFS) in this 
noisy model except for some special directions of field, like $\vec{n} = (0,0,1)^T$, etc. In addition it may happen that some quantum states are 
completely invariant for certain directions of field as well. However, an interesting and non-trivial possibility is the time-invariant entanglement 
such that the quantum states are changing at every instance however their entanglement remains constant. Such observation 
was initially made for qubit-qutrit systems \cite{Karpat-PLA375-2011} and later on for a specific family which is so called Bell-diagonal states of two 
qubits \cite{Liu-arXiv}. The time-invariant entanglement phenomenon including our current study has been observed for $\vec{n} = (0,0,1)^T$. 
This direction of field always have DFS for all dimensions of Hilbert space. The combination of quantum states residing in DFS 
and subspaces orthogonal to them lead to this phenomenon as explained below.

By choosing $\vec{n} = (0,0,1)^T$ onwards in this Letter, first we take an example of two qubits. The resulting time evolved quantum state for two 
qubits can be obtained straightforwardly as
\begin{eqnarray}
\rho(t) = \left( 
\begin{array}{cccc}
\rho_{11} & \gamma \, \rho_{12} & \gamma \, \rho_{13} & \gamma^2 \, \rho_{14} \\ 
\gamma \, \rho_{21} & \rho_{22} & \rho_{23} & \gamma \, \rho_{24} \\ 
\gamma \, \rho_{31} & \rho_{32} & \rho_{33} & \gamma \, \rho_{34} \\
\gamma^2 \, \rho_{41} & \gamma \, \rho_{42} & \gamma \, \rho_{43} & \rho_{44}
\end{array}
\right)\,,
\label{Eq:r2Qt}
\end{eqnarray}
where $\gamma = e^{-\Gamma t}$. We note that there are two sectors in which entangled states can reside. For Bell states 
$|\Phi^\pm \rangle = 1/\sqrt{2} (|00\rangle \pm |11\rangle)$, there is decay of entanglement, whereas Bell states 
$|\Psi^\pm \rangle = 1/\sqrt{2} (|01\rangle \pm |10\rangle)$ remain invariant while being in DFS. If we mix any Bell state living in 
DFS with any other Bell state in orthogonal subspace then we may find the phenomenon of time-invariant entanglement. 
As a concrete example, let us consider the family of states 
\begin{eqnarray}
\rho_{a, b} = \, b \, |\Psi^\pm\rangle\langle \Psi^\pm| + (1-b) \, \rho_a \,,
\label{Eq:Rab}
\end{eqnarray}
where $0\leq b \leq 1$ and $\rho_a$ is defined as 
\begin{eqnarray}
\rho_a = \, a \, |\Phi^+ \rangle\langle \Phi^+ | + \frac{1-a}{4} \,\mathbb{I}_4 \,,
\label{Eq:ra}
\end{eqnarray}
where $0\leq a \leq 1$. The time evolution of states $\rho_{a,b}$ can be straightforwardly written as
\begin{eqnarray}
\rho_{a, b}(t) = \, b \, |\Psi^\pm\rangle\langle \Psi^\pm| + (1-b) \, \rho_a(t) \,.
\label{Eq:Rabt}
\end{eqnarray}
The four eigenvalues of the partially transposed matrix $\rho_{a, b}^{T_A}(t)$ are $[1 + a - (3 + a) \, b ]/4$, 
$[1 + a + b - a \, b ]/4$, $[1 + b - a (1 - b)(1 - 2 e^{-2 \Gamma t}) ]/4$, and $[1 + b - a (1 - b)(1 + 2 e^{-2 \Gamma t}) ]/4$. As it is known that 
for two qubits, the partially transposed matrix can have maximum one negative eigenvalue, therefore for the choice  
\begin{eqnarray}
b > \frac{1 + a}{3 + a} \,, 
\label{Eq:cnd2qb}
\end{eqnarray}
the first eigenvalue is negative and the rest of the three eigenvalues are positive. As this negative eigenvalue is time-invariant, therefore 
we can quantify entanglement by negativity as 
\begin{equation}
 N(\rho_{a, b}(t)) = \frac{(3 + a) b - 1 - a}{2}\,,
\end{equation}
causing time-invariant entanglement although the quantum states are changing at every instance as evident by their eigenvalues. 
If condition~(\ref{Eq:cnd2qb}) is not satisfied then entanglement decays. As a special case of 
$a = 1$, $b = 0.7$ and $b = 0.75$, we recover the results already worked out recently \cite{Liu-arXiv}. 

\begin{figure}
\scalebox{2.00}{\includegraphics[width=1.95in]{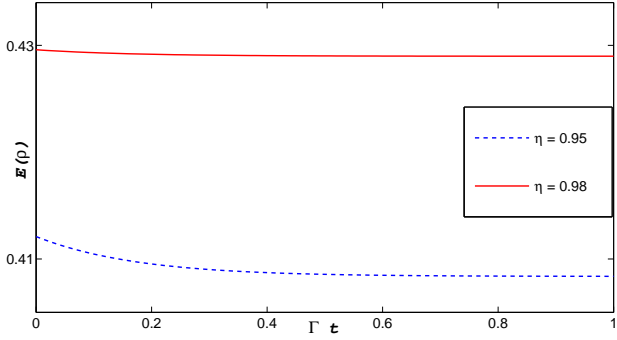}}
\caption{Genuine negativity for three qubits is plotted against parameter $\Gamma t$ for $\rho_\eta(t)$ states (Eq.~(\ref{Eq:eta}))
for two values of parameter $\eta$. See text for details.}
\label{FIG:EFGW1}
\end{figure}

Let us consider three qubits. In this case, we have two types of inequivalent genuinely entangled states, namely $GHZ$ type states and $W$ type state. 
The most general solution for an arbitrary initial state is similar to Eq.~(\ref{Eq:r2Qt}), such that $W$ state and $\tilde{W}$ 
(locally equivalent state) reside in decoherence free subspace. The natural extension of Bell-diagonal states for multi qubits are GHZ-diagonal states. 
In computational basis these states lie on the main diagonal and anti-diagonal of density matrix, hence forming an X just like Bell-diagonal states.   
The GHZ-diagonal states are subset of X-states and all off-diagonal matrix elements decay under collective dephasing. Therefore in order to look for 
time-invariant entanglement we must take mixture of GHZ states and W states. The only GHZ state $|GHZ \rangle = (|000\rangle \pm |111\rangle)/\sqrt{2}$ 
has no overlap with W state, so we define our quantum states as
\begin{equation}
\rho_\eta = (1 - \eta) \, |GHZ \rangle \langle GHZ | + \eta \, |W \rangle \langle W | \,,  
\end{equation}
where $0 \leq \eta \leq 1$ and $|W\rangle = 1/\sqrt{3} (|001\rangle + |010\rangle + |100\rangle)$. The time evolved states are written as
\begin{equation}
\rho_\eta (t)= (1 - \eta) \, |GHZ(t) \rangle \langle GHZ(t) | + \eta \, |W \rangle \langle W | \,.  
\label{Eq:eta}
\end{equation}
As it is hard to find analytical expressions for genuine negativity, nevertheless, based on our numerical search, we conjecture that 
there is no time-invariant genuine entanglement for three qubits under collective dephasing.  
Figure (\ref{FIG:EFGW1}) shows the genuine negativity for two values of parameter $\eta$. We see that although entanglement is changing at a very 
slow rate due to very large percentage of $W$ state, nevertheless, we do not have any time-invariant entanglement. Our numerical calculations suggests that 
we do not have any time-invariant entanglement even for $\eta = 0.99$.

\begin{figure}
\scalebox{2.00}{\includegraphics[width=1.95in]{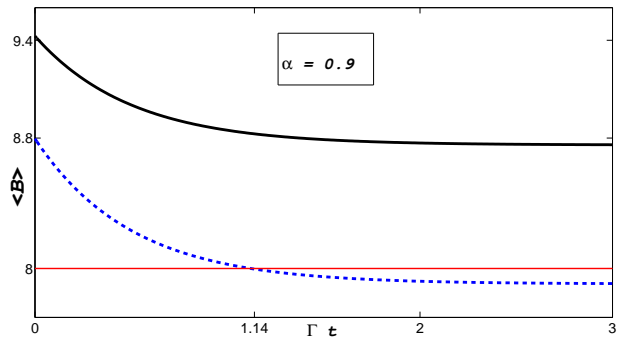}}
\caption{Genuine nonlocality is plotted against parameter $\Gamma \, t$ for states $\rho_{\alpha,\beta}(t)$. We take $\alpha = 0.9$. Top curve is for 
$\beta = 0.85$ and depicts asymptotic nonlocal states. The lower curve is for $\beta = 0.8$ and shows sudden death of genuine nonlocality.}
\label{FIG:GNL1}
\end{figure}

Finally we move to four qubits case. We demonstrate explicitly that time-invariant entanglement can occur for this dimension of 
Hilbert space.  The sixteen GHZ states in this case are defined as
\begin{equation}
 |GHZ_i \rangle = \frac{|x_1 x_2 x_3 x_4\rangle \pm |\bar{x}_1 \bar{x}_2 \bar{x}_3 \bar{x}_4\rangle}{\sqrt{2}}\,,
\label{Eq:GHZD} 
\end{equation}
where $x_j,\bar{x}_j \in \{0,1\}$ and $x_j \neq \bar{x}_j$. We note that in contrast to the three qubits, four qubits case do have some GHZ states living 
in DFS. More specifically, all those GHZ states which have two $1$'s in a ket reside in DFS, for an example, the state 
$(|0011\rangle \pm |1100\rangle)/\sqrt{2}$, and other states with permutations. We are now in a position to define a family of quantum 
states similar to two qubits case as 
\begin{equation}
\rho_\alpha = \alpha \, |GHZ_2\rangle\langle GHZ_2 | + \frac{1-\alpha}{16} \, \mathbb{I}_{16} \,, 
\label{Eq:raph}
\end{equation}
where $|GHZ_2\rangle = (|0001\rangle + |1110\rangle)/\sqrt{2}$ and $0 \leq \alpha \leq 1$. We take a mixture of this state with a state which resides in DFS, 
given as
\begin{equation}
\rho_{\alpha,\beta} = \beta \,  |GHZ_6\rangle\langle GHZ_6 | + (1-\beta) \, \rho_\alpha \,,
\end{equation}
where $|GHZ_6\rangle = (|0101\rangle + |1010\rangle)/\sqrt{2}$ and $0 \leq \beta \leq 1$. The time evolution of these states can be written as
\begin{equation}
\rho_{\alpha,\beta}(t) = \beta \,  |GHZ_6\rangle\langle GHZ_6 | + (1-\beta) \, \rho_\alpha (t)\,.
\label{Eq:abt}
\end{equation}
We can analyze the eigenvalues of the partially transposed matrix $\rho_{\alpha,\beta}^{T_A}(t)$, however we should remember the fact that there are 
states which are NPT under each partition nevertheless they are biseparable \cite{Sabin-EPJD-2008}. 
Although Figure (\ref{FIG:GNL1}) can be regarded as manisfestation of explicit time dependence of quantum states, nevertheless, the spectrum of 
quantum states $\rho_{\alpha,\beta}(t)$ can also show their explicit time dependence. The eigenvalues of these states are 
$(1-\alpha)(1-\beta)/16$ (13 times), $(1 - \alpha + 15 \, \beta + \alpha \, \beta)/16$, $(1 + 7 \, \alpha - 8 \, \alpha \, e^{-2 \, \Gamma t})(1-\beta)/16$, 
and $(1 + 7 \, \alpha + 8 \, \alpha \, e^{-2 \, \Gamma t})(1-\beta)/16$. The last two eigenvalues are time dependent and exhibit the fact that 
quantum states are changing at all times.

\begin{figure}
\scalebox{2.00}{\includegraphics[width=1.95in]{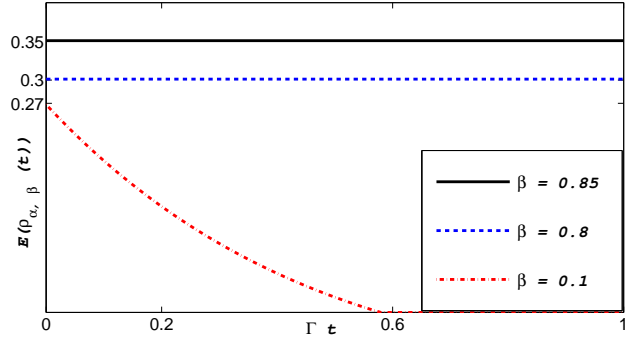}}
\caption{Genuine negativity for four qubits is plotted against parameter $\Gamma t$ for $\rho_{\alpha,\beta}(t)$ states (Eq.~(\ref{Eq:abt})) 
for various values of parameter $\beta$. We take $\alpha = 0.9$. See text for details.}
\label{FIG:GHZEF1}
\end{figure}

Figure \ref{FIG:GHZEF1} shows genuine negativity plotted against parameter $\Gamma \, t$ for family of states $\rho_{\alpha,\beta}(t)$. We have set 
$\alpha = 0.9$ and plotted three instances for parameter $\beta$. We can see that for $\beta = 0.85$ and $\beta = 0.8$, we have time-invariant 
genuine entanglement, whereas for $\beta = 0.1$, we have decay of entanglement. For smaller values of $\beta$, the subspace orthogonal to DFS is  
dominant and entanglement decays, whereas for larger values of $\beta$, we may get time-invariant entanglement even though the quantum 
states are changing at every instance. Hence we have explicitly demonstrated the existance of time-invariant feature for genuine entanglement of four qubits. 

In order to study the genuine nonlocality of states $\rho_{\alpha,\beta}(t)$, we need to find appropriate measurement operators for it. As time-invariant 
entanglement occurs for larger values of $\beta$, which implies that the largest off-diagonal matrix element corresponds to state $|GHZ_6\rangle$. As all 
GHZ states are locally equivalent to each other via a local unitary operator, therefore we can apply same local unitary transformations to measurement 
operators and then take expectation value of Bell operator $\mathcal{B}_A$. The expectation value of such operator is given as  
\begin{eqnarray}
\langle \mathcal{B}_A \rangle =  \frac{16 \, \beta - \alpha \, (1-\beta) \, (9 - 7 \, e^{-2 \, \Gamma t})}{\sqrt{2}}\,. 
\end{eqnarray}
We note that for $\beta = 1$, we have maximum violation of $8 \sqrt{2}$, which is expected as the state is in DFS and not changing so its genuine nonlocality 
remains invariant. Whereas for $\beta < 1$, we have decay of genuine nonlocality provided that $\alpha \neq 0$. Depending upon parameter $\beta$, we can 
either have nonlocal states throughout the dynamics or sudden death of genuine nonlocality. We find that for  
\begin{equation}
 \beta > \frac{4 \sqrt{2} + \alpha}{8 + \alpha} \ , 
\end{equation}
we have initial genuine nonlocal quantum states. It is well known that the $n$-partite quantum state $\rho(t)$ exibits genuine multipartite nonlocality if 
$|\langle \mathcal{B} \rangle| > 2^{n-1}$. In Figure (\ref{FIG:GNL1}), we plot the expectation value of $\mathcal{B}_A$ against parameter $\Gamma \, t$ for 
$\beta = 0.85$ (black solid line) and $\beta = 0.8$ (blue dashed line). We see that we have asymptotic nonlocal states for top curve and sudden death of 
nonlocality for lower curve. In both situations, these curves indicate the fact that quatum states are changing at every instance, such that genuine 
nonlocality is changing whereas its genuine entanglement does not change at all. 

In summary we have studied the dynamics of genuine entanglement and genuine nonlocality for multiqubits quantum systems under collective dephasing. 
We have investigated the possibility of time-invariant entanglement for two, three and four qubits. We have found that there exist a non-trivial feature 
of quantum states, in which the states are changing at every instance whereas their entanglement remain constant. The change in quantum states 
might be indicated by the change in their nonlocality. This feature was first discovered for qubit-qutrit systems and later on for qubit-qubit quantum 
states. Based on our numerical search, we conjecture that there is no evidence of time-invariant entanglement for three qubits states. 
For four qubits, we have explicitly demonstrated phenomenon of time-invariant genuine entanglement for a family of quantum states. It seems that 
if we have mixtures of two entangled states such that one entangled state lives in decoherence 
free subspace and other entangled state resides in orthogonal subspaces to it, only then we might observe this phenomenon. The fraction of entangled states  
living in DFS must be larger than entangled states in orthogonal subspaces. We have also studied the decoherence effects on genuine nonlocality of 
those quantum states which exhibit time-invariant entanglement. We have found that these states may exhibit either asymptotic nonlocality or so called 
sudden death of nonlocality, in which quantum states loose their nonlocality at a finite time. Although the quantum states remain genuine entangled, 
nevertheless, their genuine nonlocality varies. We believe that our findings can be verified in already available ion trap or photonic experiments, 
for example, in $^{40}Ca^+$ ion, the operation of $MS_\phi(\pi/2)$ can map the ground state $|0000\rangle$ of four qubits directly to 
maximally entangled GHZ state \cite{Schindler-NJP2013}. Any other GHZ state can be obtained by rotation $R_\phi(\pi/2)$. 
The state $\rho_\alpha$ (Eq.~(\ref{Eq:raph})) can be prepared via same technique as Werner states have been prepared with operation 
$MS2 = \exp(- i \pi/4 \sigma_J \sigma_J)$ using 854nm and 729nm pulses \cite{Lanyon-PRL111-2013} and so on. Experimental verification of 
time-invariant entanglement of two qubits \cite{Liu-arXiv} indicates the interest and importance of such experiments for multipartite quantum systems. 
Finally, one may ask how the results obtained here may change if we have local decoherence instead of global. As we have pointed out that the presence of 
decoherence free subspaces seems to be a necessary condition for the time-invariant entanglement and it is known that for local dephasing there are no
decoherence free subspaces, therefore, we expect no such phenomenon in that case. Our previous studies on dynamics of genuine entanglement for local 
decoherence also support this claim \cite{Ali-JPB-2014}. One of the future avenues would be to look for the time-invariance of quantum nonlocality.  

\acknowledgments
M. Ali is grateful to Edoardo G. Carnio for helpful discussions and Otfried G\"uhne for his correspondence. The author is also thankful to both 
referees for their constructive and helpful comments which brought much clarity in the Letter. 

%

\end{document}